%% file: manuscript.tex
\documentclass[conference]{IEEEtran}
\IEEEoverridecommandlockouts
\usepackage[left=1.59cm,right=1.58cm,top=1.78cm]{geometry}
\usepackage{caption}
\usepackage{subcaption}
\usepackage{cite}
\usepackage{times}
\usepackage{amsmath}
\usepackage{algorithm}
\usepackage[noend]{algorithmic}
\usepackage{stfloats}
\usepackage{caption}
\usepackage{color}
\usepackage{fancyvrb}
\usepackage{amsmath}
\usepackage{amssymb}
\usepackage{makeidx}
\usepackage{mdwmath}
\usepackage{mdwtab}
\usepackage{moreverb}
\usepackage{fancyvrb}
\usepackage{etoolbox}
\usepackage{mathtools}
\usepackage{mathptmx}
\usepackage{multirow}
\usepackage{amsmath,amssymb,amsfonts}
\usepackage{algorithmic}
\usepackage{graphicx}
\usepackage{authblk}
\usepackage{url}

\usepackage{textcomp}
\usepackage[nolist]{acronym}

\usepackage{tikz}

\usepackage{xcolor}
\def\BibTeX{{\rm B\kern-.05em{\sc i\kern-.025em b}\kern-.08em
 T\kern-.1667em\lower.7ex\hbox{E}\kern-.125emX}}

\begin{document}

\title{
    \vspace{-0cm} 
    \begin{tikzpicture}[remember picture, overlay]
        \node[anchor=north, yshift=-0.5cm] at (current page.north) {\fbox{\parbox{\textwidth}{\centering\small {\color{red}This version of the paper has been accepted for presentation at IEEE CPSCom 2023. It is subject to IEEE copyright and may be removed upon request.}}}};
    \end{tikzpicture}
    \vspace{0cm} 
    Conflict Management in the Near-RT-RIC of Open RAN: A Game Theoretic Approach

}

\vspace{-10pt}

\author[1, 3]{Abdul Wadud}
\author[1]{Fatemeh Golpayegani}
\author[1,2]{Nima Afraz}
\vspace{-5pt}
\affil[1]{School of Computer Science, University College Dublin, Ireland}
\affil[2]{UCD Beijing-Dublin International College (BDIC)}
\affil[3]{Bangladesh Institute of Governance and Management}
\affil[ ]{\textit{abdul.wadud@ucdconnect.ie} \rm{and} \textit{\{ fatemeh.golpayegani, nima.afraz\}@ucd.ie}}


\vspace{-50pt}

\maketitle

\begin{abstract}
Open \ac{RAN} was introduced recently to incorporate intelligence and openness into the upcoming generation of \ac{RAN}. Open \ac{RAN} offers standardized interfaces and the capacity to accommodate network applications from external vendors through \acp{xApp}, which enhance network management flexibility. The \ac{Near-RT-RIC} employs specialized and intelligent \acp{xApp} for achieving time-critical optimization objectives, but conflicts may arise due to different vendors' \acp{xApp} modifying the same parameters or indirectly affecting each others' performance. A standardized \ac{CMS} is absent in most of the popular Open RAN architectures including the most prominent O-RAN Alliance architecture. To address this, we propose a \ac{CMS} with independent controllers for conflict detection and mitigation between \acp{xApp} in the \ac{Near-RT-RIC}. We utilize cooperative bargain game theory, including \ac{NSWF} and the \ac{EG} solution, to find optimal configurations for conflicting parameters. Experimental results demonstrate the effectiveness of the proposed \ac{CMC} in balancing conflicting parameters and mitigating adverse impacts in the \ac{Near-RT-RIC} on a theoretical example scenario.
\end{abstract}

\begin{IEEEkeywords}
Open \ac{RAN}, \ac{Near-RT-RIC}, conflict mitigation, game theory
\end{IEEEkeywords}
\input{Acronyms}

\section{Introduction}

\IEEEPARstart{O}{\uppercase{pen}} \acp{RAN} are one of the most prominent architectures for 5G and beyond communication networks. They have been widely studied in recent years by experts from both academia and industry \cite{b1, b2}. The concept of Open \ac{RAN} has emerged with the advent of the 5G and \ac{IoT} based technologies like self-driving cars, high-speed virtual reality-based gaming, and smart factories; this is because the capabilities of the existing \ac{RAN} become inadequate to fulfil the future \ac{5G}/\ac{6G} networks and \ac{IoT} requirements \cite{b1}, including network flexibility, ultra-low latency and massively diverse connectivity. 
 {The main motivation behind introducing Open \acp{RAN} is the limitation of existing \acp{RAN} due to the monolithic nature of their components, which makes them a black box to the network operators \cite{b2}.} This black box provokes various disadvantages, including limited reconfigurability, restricted coordination among network nodes, and vendor lock-in at the operators' end. These challenges lead to the inefficient utilization of spectrum resources and cause difficulties in radio spectrum management and optimization using real-time adaptation. Open \ac{RAN} can be used to overcome these challenges, and it can provide network control to the operators using a centralized \ac{SD-RAN}. Also, Open \ac{RAN} promotes a disaggregated and virtualized architecture, connected through open and standardized interfaces and interoperable with different vendors. Promising open, resilient, reconfigurable, cost and energy-efficient \acp{RAN} that are {amenable} to data-driven closed-loop control using software-defined protocol stacks. Besides, Open \ac{RAN} extends the opportunity of using \ac{AI}/\ac{ML} to automate network management and maintenance. 

\par The \ac{RIC} maintains the network control operations in open \ac{RAN}, enabling various vendors to install control applications that manage specific network targets (e.g., resource allocation, energy saving, mobility load balancing, and more {\cite{b2}}). Non-time critical tasks that can be completed in more than $1s$ are performed by \acp{rApp} installed in the \ac{Non-RT-RIC}, whereas time-critical tasks that should be completed within $10ms$ to $\leq 1s$ are performed by \acp{xApp} installed in the \ac{Near-RT-RIC}. 
These \acp{xApp} and \acp{rApp} control the network operation and management within the \ac{RIC}. For instance, an \ac{rApp} can improve network performance and reduce latency during a \ac{V2X} dynamic handover scenario by adjusting radio resource allocation policies \cite{rimedo_labs}. Similarly, an \ac{xApp} can maximize \ac{QoS} for a block of users by managing radio resources and sending dedicated control messages to the \ac{RAN} nodes \cite{b3}. As these \acp{xApp} and \acp{rApp} can be provided by various vendors and they share the same resources concurrently during network operation, they may influence each other's performance negatively \cite{b4}. These interferences in control decisions, known as conflicts, must be detected and mitigated; otherwise, they can result in significant performance degradation within the system. 

\par In this study, we discuss the taxonomy of possible conflicts within the Open \ac{RAN} architecture. We propose a \ac{CMS} that can detect various types of conflicts in the \ac{Near-RT-RIC} among \acp{xApp} and mitigate them based on iterative game theoretic approaches and Nash equilibrium. The proposed \ac{CMS} has an independent \ac{CDC} that detects conflicts using the similar architecture proposed in \cite{b6}, but it functions entirely uniquely (see section~\ref{cdc}). The proposed \ac{CMS} in this paper uses a \ac{QoS} threshold to trigger the \ac{CDC} that ensures guaranteed minimum service requirement in Open \ac{RAN}. The \ac{CMC} component of the proposed \ac{CMS} uses \ac{NSWF} or \ac{EG} solution to maximize the collective performance of the network. We use an experimental model in section~\ref{sec:num_res} to analyze the performance of the proposed \ac{CMC} and discuss the functionalities of the proposed solution. Although we only deal with conflicts between \acp{xApp} in the \ac{Near-RT-RIC} in this article, our proposed solution can be extended to mitigate all types of conflicts in Open \ac{RAN} with proper infrastructure and planning. 

\par The remainder of this paper is structured as follows. In section~\ref{sec:arc}, we briefly discuss the architecture of Open \ac{RAN} with all the significant components, splits, and interfaces. Section~\ref{sec:con_OpenRAN} presents a taxonomy of all possible conflicts in Open RAN with proper definitions and instances. The required infrastructure for the proposed \ac{CMS} is discussed in section~\ref{sec:infra_db}. In section~\ref{sec:CMS}, we discuss the proposed \ac{CMS} in detail. A theoretical example model is used to evaluate the performance of the proposed \ac{CMS}; numerical results and analysis in this regard are discussed in section~\ref{sec:num_res}. Challenges and limitations of the proposed \ac{CMS} are discussed in section~\ref{sec:limit}. Finally, we conclude the paper in section~\ref{sec:con_fut} with concluding remarks and future research direction.


\section{Open \ac{RAN} Architecture}
\label{sec:arc}
\par
\begin{figure}[!ht]
 \centering
 \vspace{-0.2in}
	\includegraphics[scale=0.5]{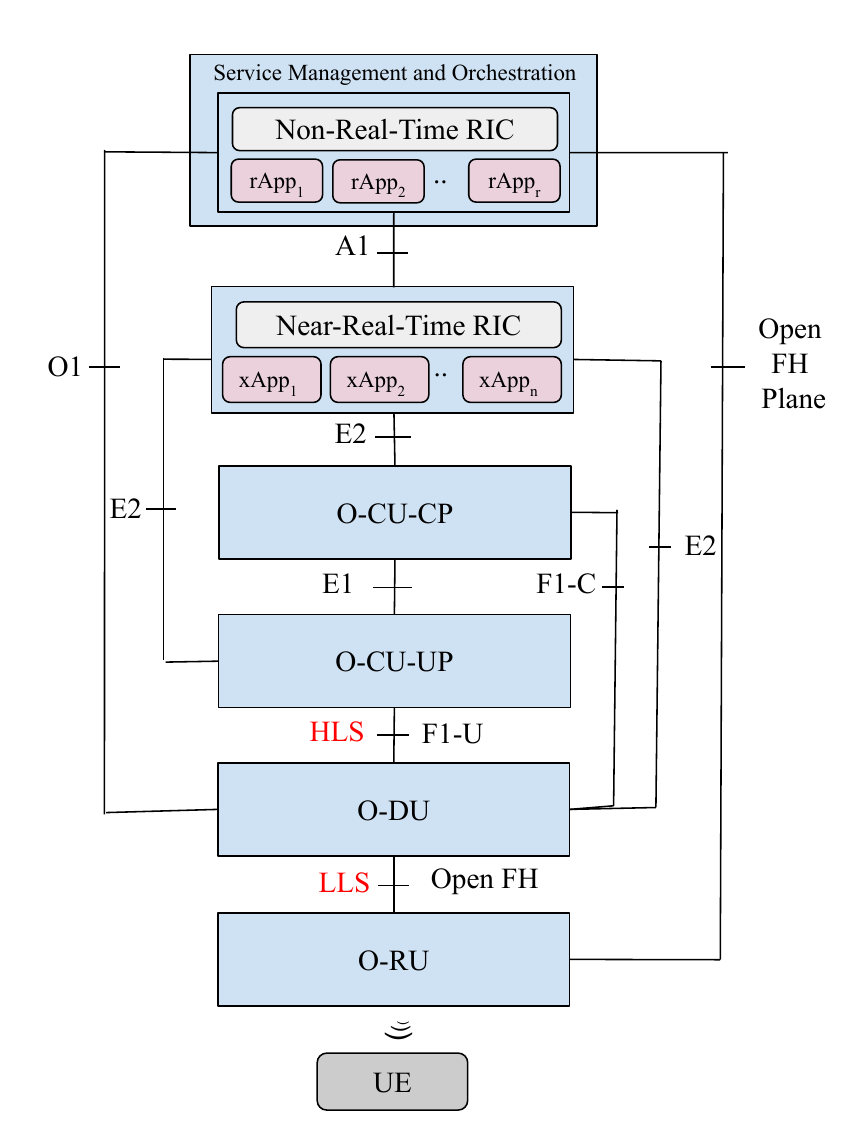}
	\vspace{-0.1in}
	\caption{Architecture of Open \ac{RAN}}
	\label{fig:arc}
\vspace{-0.2in}
\end{figure}

 The traditional \ac{RAN} base station is segregated into three major units (see in Fig.~\ref{fig:arc}), namely- open centralized unit (O-CU), open distributed unit (O-DU), and open radio unit (O-RU), which is also known as remote radio units (RRU). The O-CU is further divided into control planes and user planes. The CUs handle higher-layer protocols and larger time-scale functions. In contrast, the DUs control lower-layer protocols and time-critical tasks, and the RRUs/RUs deal with physical layers and manage \ac{RF} communications and \ac{RF} components. The open interface between the RU and DU is called the lower-layer splits (LLS), whereas the interface between DU and CU is called the higher-layer splits (HLS). 

\par In the HLS, the F1 interface is used to connect the DUs with the CUs with several other dedicated sub-interfaces for user and control planes. In the LLS, the front haul control-user-synchronization (CUS) and management plane interfaces are used to connect RUs with DUs. Each interface is defined with a combination of well-organized procedures and plays a vital role in the Open \ac{RAN} ecosystem by enabling different types of control and automation to the network. All required applications for these interfaces run on the \ac{RIC}. The \ac{RIC} is one of the significant elements for enabling various key 5G features, including network slicing, low-latency applications, prioritized communications, high bandwidth, etc.

\par In the architecture Open \ac{RAN}, the \ac{RIC} is categorized as \ac{Non-RT-RIC} and \ac{Near-RT-RIC}. The former deals with events and resources with a response time of 1 second or higher, whereas the latter manages events and resources with a response time on a scale of 10 milliseconds to 1 second. The \ac{Non-RT-RIC} operates on the operators’ network centrally, whereas the \ac{Near-RT-RIC} can operate both centrally or on the edge of the network. A set of standardized interfaces, including A1, E1, E2, F1, O1, Open \ac{RAN} front haul, and others, are used in different parts of the Open \ac{RAN} architecture and enable the \ac{Non-RT-RIC} and \ac{Near-RT-RIC} to support the control loops \cite{b2}. As a whole, the architecture of Open \ac{RAN} supports its foundational principles of openness, disaggregation, virtualization, data-driven control, and open interfaces. 

\par Although the proposed Open \ac{RAN} architecture in Fig.~\ref{fig:arc} by O-RAN Alliance (an organization of \acp{MNO}) \cite{b5} is capable of handling challenges of the existing \ac{RAN} architecture, it brings a few new and unique challenges. The conflict between Open \ac{RAN} components is one of such challenges. Various types of conflicts among Open \ac{RAN} components are discussed in the following section.

\section{Conflicts in Open \ac{RAN}}
\label{sec:con_OpenRAN}
In traditional \ac{RAN} with a single vendor, conflicts were typically resolved internally by the vendor within their equipment. Since the entire \ac{RAN} infrastructure was provided by a single vendor, they had control over the design, configuration, and optimization of the network. Any conflicts or compatibility issues were handled within their closed ecosystem, resulting in a more streamlined prevention and resolution process. However, with the introduction of Open \ac{RAN}, the network architecture allows the integration of equipment and software components from multiple vendors. While this promotes interoperability and flexibility, it also introduces potential conflicts between the various components. Each vendor may have different approaches, optimizations, or parameter configurations, which can lead to clashes when integrating their solutions within the Open \ac{RAN} framework and negatively influence the \ac{RAN}'s performance. Therefore, these conflicts should be mitigated or avoided with proper network configuration \cite{b2}. Types of various control conflicts are discussed in the following subsections.

\subsection{Types of Conflicts in Open \ac{RAN}}
\label{types_conf}
\begin{figure}[!ht]
 \centering
 \vspace{-0.1in}
	\includegraphics[scale = 0.42]{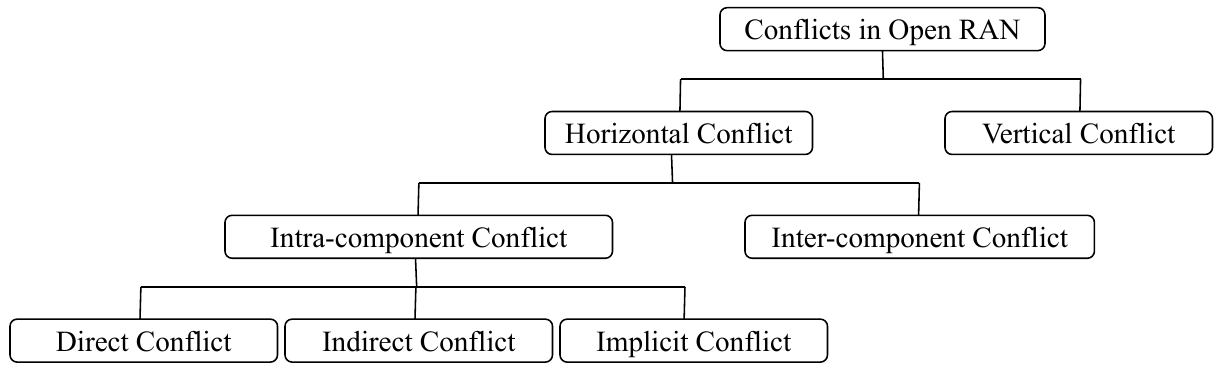}
	\vspace{-0.1in}
	\caption{Taxonomy of potential conflicts in Open \ac{RAN}.}
	\label{fig:tax}
 \vspace{-0.1in}
\end{figure}

According to Adamczyk et al. in \cite{b6}, control decision conflicts may occur in different levels of Open \ac{RAN} architecture. Conflicts in Open \ac{RAN} are primarily classified into horizontal and vertical conflicts (see Fig.~\ref{fig:tax}). 

\par
\begin{figure}[!ht]
 \centering
 \vspace{-0.2in}
	\includegraphics[scale=0.6]{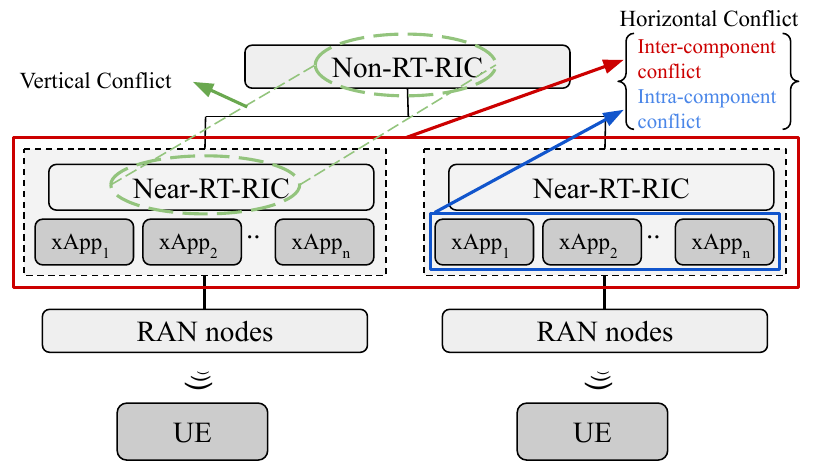}
	\vspace{-0.1in}
	\caption{Potential conflicting areas in Open \ac{RAN}.}
	\label{fig:area_conf}
 \vspace{-0.1in}
\end{figure}

 Vertical conflict occurs between components of different levels in the Open \ac{RAN} architectural hierarchy. For example, the control decision conflict between a \ac{Near-RT-RIC} and a \ac{Non-RT-RIC} illustrated in Fig.~\ref{fig:area_conf} is termed a vertical conflict. Horizontal conflict, on the other hand, happens between components of the same level. For example, a conflict between two \acp{xApp} in the \ac{Near-RT-RIC} or between two neighboring \acp{Near-RT-RIC} is considered a horizontal conflict (see Fig.~\ref{fig:area_conf}). Control-decision conflict among \acp{xApp} within a \ac{Near-RT-RIC} is referred to as intra-component conflict, and conflict among \acp{xApp} of neighboring \acp{Near-RT-RIC} is referred to as inter-component conflict (see Fig.~\ref{fig:area_conf}). Intra-component conflict can be further categorized into direct, indirect, and implicit conflict. In this article, we propose a mitigation technique for intra-component conflict between \acp{xApp} in the \ac{Near-RT-RIC}.

\par In the \ac{Near-RT-RIC}, independent \acp{xApp} functioning to achieve different optimization goals may result in conflicting configurations by modifying the same parameter or indirectly influencing the same parameter \cite{b4}; we categorize these conflicts as intra-component conflict. Detecting these direct, indirect, and implicit conflicts is difficult as the involved \acp{xApp} might be developed and supplied by different vendors and are unlikely to exchange information among each other \cite{b7}. 

\par Direct conflict can be detected directly and handled by the internal conflict mitigation controller. A single \acp{xApp} or two \acp{xApp} may request conflicting configurations with the current running configuration. Also, two or more \acp{xApp} may ask to change a particular metric to different values, which leads to a direct conflict. In both scenarios, the conflict mitigation controller processes these requests and decides which demand prevails in the action. This type of resolution is called the pre-action resolution \cite{b4}. However, prioritizing one request over the other is not considered an optimal solution for this problem. A more appropriate approach involves identifying an optimal configuration that addresses conflicting metrics or parameters. This not only ensures fairness but also serves the collective interest of the network to achieve its optimization goals.


\par On the contrary, indirect and implicit conflicts cannot be observed directly. When a change in one \ac{xApp}'s parameter setting affects or influences the same area of other \acp{xApp} parameter changes, indirect conflict occurs. For example, cell individual offset (CIO) and antenna tilts are different control \acp{xApp}, but both can affect the handover boundary. When one \acp{xApp} controlling the remote electrical tilts (RET) or antenna tilts changes its values, the handover boundary in the cell is affected. This adjustment indirectly impacts the performance of the CIO controlling \acp{xApp}. Mitigating this conflict requires post-action measurement by deciding an optimal value for the conflicting parameter \cite{b4}. 

\par Similarly, implicit conflict transpires when two \acp{xApp} independently optimize their target but cause adversarial or unwanted effects on each other's performance. For example, two \acp{xApp} maximizing \ac{QoS} for a group of users and minimizing the number of handovers, respectively, may disrupt each others' performance by interfering in a non-obvious manner. This conflict is the most difficult to detect and mitigate. However, a generic approach of careful resource planning, prioritization, and error handling can help ensure the system's smooth operation \cite{b4}. In this article, we propose a conflict management system that can be used to successfully detect and mitigate all the types of conflicts, as mentioned earlier, among \acp{xApp} in a \ac{Near-RT-RIC}.

\section{Infrastructure Prerequisites and Database}\label{sec:infra_db}
The proposed \ac{CMS} demands additional components and configurations in the Open \ac{RAN} architecture. Two dedicated messaging channels in the messaging infrastructure need to be configured from \acp{xApp} to the \ac{Near-RT-RIC}'s centralized database and the conflict mitigation controller, respectively (see Fig.~\ref{fig:CMS_arc}). The former redirects all control messages from \acp{xApp} to the centralized database, and the latter acts as a medium between all \acp{xApp} and the \ac{CMC} to find the optimal configuration of the conflicting parameter(s). Iterative communication during the optimal value calculation by the \ac{CMC} with \acp{xApp} and pipeline for processing all messages in the messaging infrastructure adds additional computational complexity in the \ac{Near-RT-RIC}, which should be acknowledged before adopting this solution. The components of the \ac{Near-RT-RIC}'s database are briefly discussed in the following.

\subsection{Database}
\label{database}
\subsubsection{\acp{RCP}}
All recently changed parameters by the request of various \acp{xApp} are stored under \ac{RCP} component of the database with their respective timestamps. 

\subsubsection{\ac{PGD}}
This component of the database stores parameters that influence the same area in the network. For example, antenna tilts and cell individual offset affect the handover boundary of a cell; therefore, both \acp{ICP} are included in the handover boundary group. 

\subsubsection{\ac{RCPG}}
If any parameter is changed that belongs to \ac{PGD}, it is stored under \ac{RCPG} component of the database with its respective timestamp and other involved parameters of the same group.

\subsubsection{\ac{PKR}}
This component of the database stores the minimum and maximum possible value of the individual parameter and the \ac{KPI} of the particular cell.

\subsubsection{\ac{DCKD}}
This component of the database stores individual \ac{KPI} threshold based on the \ac{QoS}/\ac{SLA} requirement of respective cells/networks. 

\subsubsection{\ac{KDO}}
This component of the database stores any \ac{KPI} degradation occurrence after every parameter changes by the \acp{xApp} through \ac{RAN} nodes with respective timestamps. 

\begin{figure}[!ht]
 \centering
 \vspace{-0.1in}
	\includegraphics[scale=0.6]{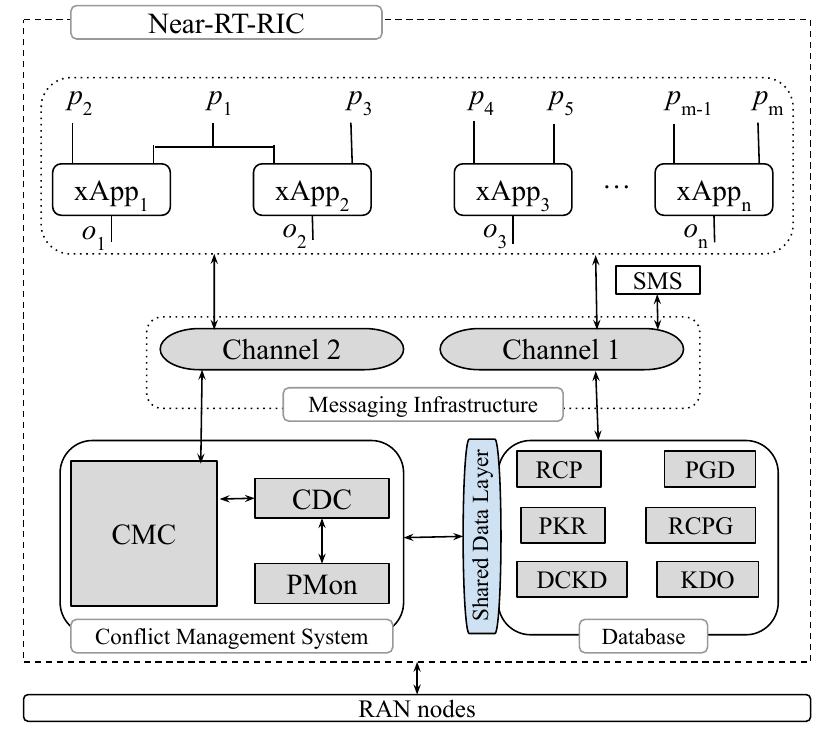}
	\vspace{-0.08in}
	\caption{Architecture of CMS in the \ac{Near-RT-RIC}.}
	\label{fig:CMS_arc}
 \vspace{-0.1in}
\end{figure}

\section{Conflict Management System}
\label{sec:CMS}
In this study, we propose a conflict management system that adopts the cooperative game theory and iterative bargaining concept to find optimal configurations for the conflicting parameters. The \ac{CMS} consists of two major components, viz- the \ac{CDC} and the \ac{CMC}. The \ac{CDC} detects conflicts with the help of a centralized database that records every action taken by an \ac{xApp} with respective timestamps. The \ac{PMon} component informs the \ac{CDC} to trigger when a specific \ac{KPI} is unable to meet the \ac{QoS} threshold. The \ac{CDC} accesses the database through a \ac{SDL} and fetches the most recently changed parameter(s) that are responsible for the recent \ac{KPI} degradation. 

\par Two states of those parameters' values are passed to the \ac{CMC} for processing an optimal configuration of the conflicting parameter(s). One state fetches the last record when the \ac{KPI} was above the \ac{QoS} threshold, and the other fetches the most recent record after the degradation transpired. The \ac{CDC} detects the exact parameter responsible for that specific \ac{KPI} degradation using the \ac{PGD} table and \ac{RCP} table and passes the above-stated two records to the \ac{CMC}. Afterwards, the \ac{CMC} communicates respective xApp(s) dealing with that specific parameter through a dedicated messaging channel and suggests an optimal configuration for the conflicting parameter using a bargain game theoretic approach. The following subsections discuss the functionalities of the proposed \ac{CMS} with appropriate figures and examples. 

\subsection{Performance Monitoring Component (\ac{PMon})}
The \ac{PMon} component in the \ac{CMS} is responsible for monitoring Key Performance Indicators (\acp{KPI}) to assess and analyze the performance of the network. The component collects data from various \ac{RAN} nodes (see in Fig.~\ref{fig:CMS_arc}) that include network elements, user devices, and other monitoring probes. This data includes measurements, statistics, and events related to network performance. The collected data is processed and analyzed to calculate the values of the defined \acp{KPI}. This involves applying algorithms and calculations to the collected data to derive meaningful performance metrics. Afterwards, the performance monitoring component compares the calculated \ac{KPI} values against predefined \ac{QoS} thresholds or target values. \ac{QoS} thresholds define acceptable ranges for each \ac{KPI}, and if a \ac{KPI} falls outside the defined \ac{QoS} range, it indicates a performance deviation or anomaly. These performance deviation occurrences are recorded in the \ac{KDO} component of the database. The \ac{PMon} component sends a trigger to the \ac{CDC} when certain \acp{KPI} breaches predefined \ac{QoS} thresholds. Afterwards, the \ac{CDC} starts looking for possible conflicts between \acp{xApp} and their respective parameters. 

\subsection{Conflict Detection Controller (\ac{CDC})}\label{cdc}
The presented \ac{CDC} in Fig.~\ref{fig:CMS_arc} is capable of detecting direct, indirect, and implicit conflicts between \acp{xApp} in the \ac{Near-RT-RIC}. However, the presented \ac{CDC} may detect all types of conflicts in Open \ac{RAN} with proper infrastructure and planning. When deploying an \acp{xApp}, direct conflict can be detected through close observation of all \acp{ICP}. However, human error may occur during the deployment, and conflicted parameter(s) can be ignored. Therefore, a detection technique is essential as part of the fail-safe mechanism to detect direct conflicts. 

\par When a \ac{KPI} degradation occurrence outside the range of \ac{QoS} threshold is detected by the \ac{PMon}, the \ac{CDC} fetches the most recently changed parameter. Afterwards, the \ac{PGD} of the database is checked for finding the affiliation of that parameter to any of the defined groups. If any affiliation is found, it is considered an indirect conflict. Otherwise, the \ac{CDC} searches all recently changed parameters in the \ac{RCP} to check whether this parameter was changed by any other or the same \acp{xApp} in the recent past. If any match is found, it is considered a direct conflict. Otherwise, it is considered an implicit conflict. Eventually, the \ac{CDC} informs the \ac{CMC} about the transpired conflict with all relevant information about the conflicting parameter. 

\subsection{Conflict Mitigation Controller (CMC)}\label{cmc}
The \ac{CMC} adopts cooperative game theory to deal with various types of intra-component conflicts in the \ac{Near-RT-RIC}. When a direct conflict is detected between \acp{xApp}, the conflicting parameter's new value is suggested by the \ac{CMC} in such a way that the collective utility of all involved \acp{xApp} is maximum. By utility, we refer to the \acp{KPI} of each \ac{xApp}. The collective utility is calculated using the Social Welfare Function (SWF) discussed in section~\ref{game_th}. Similarly, an optimal value of the conflicting parameter is estimated using the SWF when an indirect or implicit conflict is detected for a recently changed parameter. The default value or the second last changed value and their respective \acp{KPI} are considered with the most recently changed value of the conflicting parameter and relevant \ac{KPI} for the optimal value calculation. The most recently changed parameter refers to the parameter that is responsible for \ac{KPI} degradation.

\begin{figure}[!ht]
 \centering
	\includegraphics[scale=0.6]{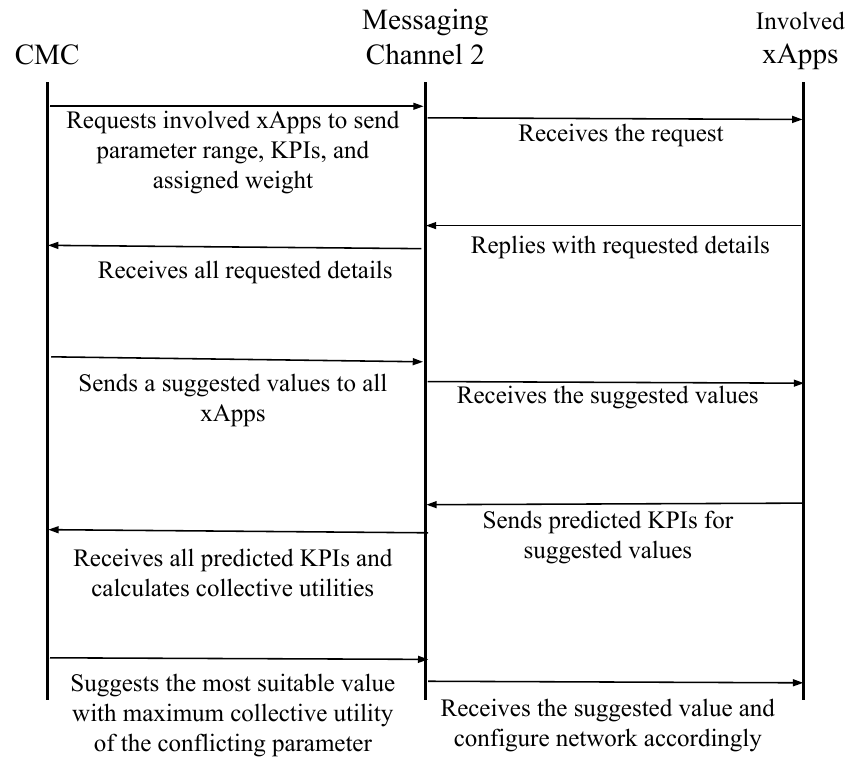}
	\vspace{-0.05in}
	\caption{Closed-loop communication between the \ac{CMC} and involved \acp{xApp} through the dedicated Channel~2 during optimal value calculation of the conflicting parameter.}
	\label{fig:cmc_com}
\end{figure}

\subsection{Theoretical Model, Assumptions and Notations} \label{theory_assump}
The Open \ac{RAN} envisions applying \ac{AI}/\ac{ML} techniques to automate the network management and optimization process. To achieve this, Open \ac{RAN} can leverage the techniques and capabilities of \ac{SON}, which is a network management and optimization technique that aims to automate the network operation. \ac{SON} deploys multi-vendor \acp{SF} that are similar to \acp{xApp} and rApps in Open \ac{RAN} to achieve specific set of objectives.

\par Table~\ref{tab:notation} presents the list of notations used in the paper.

 

\begin{table}[!ht]
 \centering
 \footnotesize
 \caption{List of Notations}
 \begin{tabular}{ p{1.5cm} p{6.6cm}}
 \hline
 \textbf{Symbol} & \textbf{Description} \\
 \hline
 \multicolumn{2}{l}{\textbf{Given Parameters:}} \\
 $X$ & Set of all \acp{xApp} in the \ac{Near-RT-RIC} \\
 $P$ & Set of all \acp{ICP} belonging to the \acp{xApp} in the \ac{Near-RT-RIC} \\
 $p \in P$ & An \ac{ICP} in the \ac{Near-RT-RIC} \\
 $O$ & Set of all \acp{KPI} belonging to the \acp{xApp} \\
 $o \in O$ & A \ac{KPI} to measure the performance of \acp{xApp} in the \ac{Near-RT-RIC} \\
 $Z$ & Set of all conflicting \acp{xApp} \\
 $J$ & Set of all \acp{KPI} for an individual \ac{xApp} \\
 $u$ & Utility of an \ac{xApp} converted from \acp{KPI} belonging to $J$ \\
 $f(x)$ & Utility function to convert \acp{KPI} to utility $u$ for a given value $x$ of the conflicting parameter \\
 $w$ & Priority weight assigned by the \ac{MNO} to a \ac{xApp} \\
 \hline
 \multicolumn{2}{l}{\textbf{Decision Variables:}} \\
 $x$ & Optimal value for the conflicting parameter \\
 \hline
 \end{tabular}
 \label{tab:notation}
\end{table}

\par We assume each \acp{xApp} similar to a \ac{SF} in \ac{SON} which carries four distinct properties. Firstly, each \acp{xApp} is capable of learning, predicting, and deciding its best possible action by itself based on any dynamic network state $s$. Secondly, \acp{xApp} are independent and do not communicate with each other. Thirdly, \acp{xApp} are provided by multiple vendors, and there are no team or group actions by any two or more \acp{xApp}. Lastly, all \acp{xApp} share the same resources and have definite conflicts of interest among them. In addition, individual \acp{xApp} can predict their respective \acp{KPI} based on all of their \ac{ICP}'s values. 

\par We further assume that an \acp{xApp} sends the following information in response to a request by the \ac{CMC} through the dedicated messaging channel (channel 2 in Fig.~\ref{fig:CMS_arc}). The first message from a \acp{xApp} to the \ac{CMC} contains the \acp{KPI} (output parameter of the \acp{xApp}) for the current value of the conflicting parameter, ranges of the conflicting parameter for a particular \acp{xApp}, and a priority weight assigned by the \ac{MNO} for a certain network state $s$. Later messages contain only the \acp{KPI} for suggested conflicting parameter's value by the \ac{CMC} (see in Fig.~\ref{fig:cmc_com}). Suppose there is $n$ number of \acp{xApp} installed in a \ac{Near-RT-RIC}. We represent the set of all \acp{xApp} as $X = \{ \rm{xApp}_1, \rm{xApp}_2, \cdots, \rm{xApp}_n \}$, set of all \acp{ICP} as $P = \{ p_1, p_2, \cdots, p_m \}$, and set of all output parameters (\acp{KPI}) as $O = \{ o_1, o_2, \cdots, o_n \}$. If there is direct conflict over a parameter $p_1$ in between \acp{xApp} $\rm{xApp}_1~\rm{and}~\rm{xApp}_2$ for a certain network state $s$, \acp{xApp} $\rm{xApp}_1$ sends $[ \{p_1^{min, \rm{xApp}_1}, p_1^{max, \rm{xApp}_1}\}, o_1^{p_1}, w_1^{s} ]$. Similarly, $\rm{xApp}_2$ sends $[ \{p_1^{min, \rm{xApp}_2}, p_1^{max, \rm{xApp}_2}\}, o_2^{p_1}, w_2^{s} ]$. Here, $w_1^{s}~\rm{and}~w_2^{s}$ are weight metrics assigned by the \ac{MNO} or chosen by an automated process to prioritize one \acp{xApp} over the other based on the necessity of the network state, and $w_1^{s} + w_2^{s} = 1$. We assume these weights as the bargaining power of \acp{xApp} $\rm{xApp}_1~\rm{and}~\rm{xApp}_2$. Set of conflicting \acp{xApp} is represented by $Z$, which is $Z = \{ \rm{xApp}_1, \rm{xApp}_2\}$ in this case. 

\par As each individual \ac{KPI} has different units and measurement ranges, the \ac{CMC} recommends converting them to a uniform scale. For instance, the \acp{KPI} can be transformed using min-max normalisation on a scale of $[0, 10]$ and defined as utility using this formula: $ f_i (x) = u_i = \frac{1}{|J|} \sum_{j \in [1, |J|]} (\frac{o_{ij}-o_{ij}^{min}}{o_{ij}^{max}-o_{ij}^{min}} \times 10))$. Here, $J$ is the set of KPIs for an individual \acp{xApp}, and $x$ is the suggested value of the conflicting parameter. For example, the performance of a power allocation \acp{xApp} can be measured using Signal-to-Interference-plus-Noise Ratio (SINR), energy efficiency, and throughput; therefore, $|J| = 3$ in this case. If there is only one KPI for evaluating the performance of an \acp{xApp}, $|J| = 1$. Values for $o_{ij}^{min}$, and $o_{ij}^{max}$ are obtained from \ac{PKR} of the database. The following section discusses how Nash's bargain game theoretic approach can be adopted for estimating the optimal value of the conflicting parameter. 

\subsection{Game Theory for Optimal Value Calculation} \label{game_th}
\begin{figure}[!ht]
 \centering
 \vspace{-0.1in}
	\includegraphics[scale=0.6]{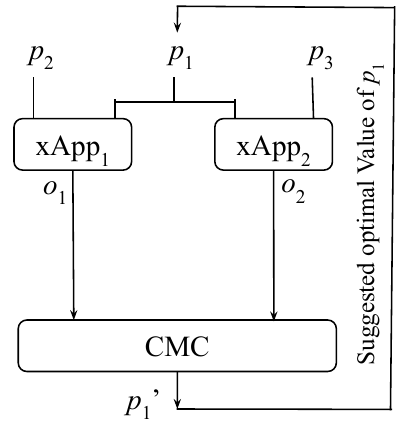}
	\vspace{-0.1in}
	\caption{Demonstration of closed-loop control for optimal value calculation by CMC.}
 \vspace{-0.15in}
	\label{fig:CMC_demo}
\end{figure}

It is well-known that Nash's equilibrium can be computed to find the maximum overall satisfaction or collective utility among all players or agents in a multi-player game or multi-agent scenario. In Fig.~\ref{fig:CMC_demo}, we consider such a conflicting scenario with two \acp{xApp} for demonstration purposes only where a direct conflict over $p_1$ between $\rm{xApp}_1$ and $\rm{xApp}_2$ is detected. When the \ac{CMC} received a trigger about this conflict, it communicates the involved \acp{xApp} immediately and requests for their \acp{KPI}, conflicting parameters range, and weights assigned by the \ac{MNO} for current network state (e.g., $o_1, \{p_1^{max, \rm{xApp}_1}, p_1^{min, \rm{xApp}_1}\}, w_1$ and $o_2, \{p_1^{max, \rm{xApp}_2}, p_1^{min, \rm{xApp}_2}\}, w_2$). The outputs are transformed into a scaled utility of similar units as discussed in section~\ref{theory_assump}. 

\par The \ac{CMC} calculates optimal range for $p_1$ considering both ranges of \acp{xApp} as- $ (p_1^{min, \rm{opt}}, p_1^{max, \rm{opt}}) \simeq \{min(p_1^{min, \rm{xApp}_1}, p_1^{min, \rm{xApp}_2}\}, max(p_1^{max, \rm{xApp}_1}, p_1^{max, \rm{xApp}_2}\})$. A sample set is prepared within this range and is used for finding the optimal value for $p_1$. For each suggested value of the sample set, collective utility is calculated using an SWF. \ac{NSWF} can calculate the collective utility using the product of the individual utilities of $n$ participating \acp{xApp}. The formula is represented as follows \cite{b8}: 

\vspace{-5pt}

\begin{equation}
NSWF(x) = \Pi_{i \in [1, |Z|]} f_{i} (x), \forall i \in Z \label{eq:nswf}
\end{equation}

\par In Eq.~\ref{eq:nswf}, \ac{NSWF} intends to balance fairness and efficiency between \acp{xApp}. However, it fails to reach optimal equilibrium when a particular \acp{xApp} requires preference over the other conflicting \acp{xApp} due to the network demand. The optimal market equilibrium, in this case, can be reached using our modified \ac{EG} convex linear programming problem as follows \cite{b9, b10}: 

\begin{subequations}
\vspace{-15pt}
\begin{eqnarray}
& & \max : F=\sum_{i \in [1, |s|]} w_i f_{i} (x) \label{eq:obj} \\
& {\rm s.t.} & \sum_{i \in [1, |Z|]} w_{i}=1, \forall i \in Z \label{eq:sum_weight} \\
& & x \geq p_m^{min, \rm{opt}}, m \in P \label{eq:min_x} \\
& &  x \leq p_m^{max, \rm{opt}}, m \in P  \label{eq:max_x} \\
& & \forall i \in [1, |Z|], |Z| \geq 2.  \label{eq:i_X}
\end{eqnarray}
\end{subequations}

Here, $F$ represents the objective function, and $x$ indicates the decision variable for obtaining the optimal configuration of the conflicting parameter. The value of $x$ is constrained within the optimal range of conflicting parameters. The following section analyses the performance of both approaches to solving various conflicts. 

\begin{figure}[!ht]
 \centering
 \vspace{-0.2in}
	\includegraphics[scale=0.6]{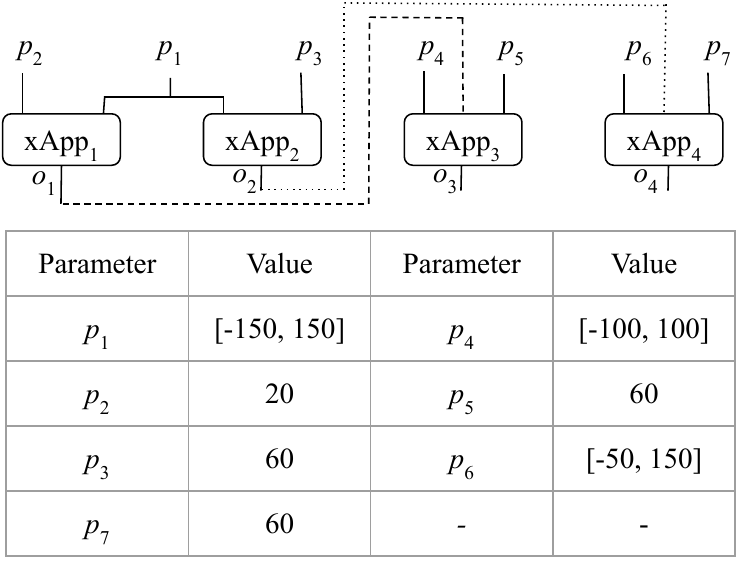}
	\vspace{-0.08in}
	\caption{Example model for direct $\{\rm{xApp_1~ and ~xApp_2}\}$, indirect $\{\rm{xApp_2~ and ~xApp_4}\}$ and implicit $\{\rm{xApp_1~ and ~xApp_3}\}$ conflict.}
	\label{fig:exam_model}
 \vspace{-0.12in}
\end{figure}

\subsection{Use cases of NSWF and EG}
\label{NSWF_EG_usecase}
Let us consider a scenario where four \acp{xApp} for Mobility Load Balancing (MLB), Capacity and Coverage optimization (CCO), Energy Saving (ES), and Mobility Robustness optimization (MRO) are functioning in a \ac{Near-RT-RIC} of Open \ac{RAN}. These \acp{xApp} aim to achieve their specific KPI objective of reducing network congestion, optimizing capacity and coverage, obtaining energy efficiency, and reducing packet/call drop rates or the number of unsuccessful handovers, respectively. Transmission power (TxP) is a common \ac{ICP} or resource that all of these \acp{xApp} share while functioning. The NSWF solution can help to suggest a value for the TxP that ensures maximum collective utility in the network. This collective utility is obtained through a trade-off between different KPIs, such as reducing energy efficiency or suffering network congestion for a certain period to reduce the number of unsuccessful handovers or call drop rates. 

\par Suppose the MNO is experiencing a high level of call drop rate in a particular cell or in a handover boundary. The Eisenberg-Galle solution can be adopted by the MNO to prioritize the MRO \acp{xApp} over the other xApps. An increment in the transmission power (TxP) will lead to higher throughput in that particular cell, or an increment in the cell's individual offset value to broaden the handover boundary can minimize the number of call drops.

\begin{figure*}
  \centering
  \begin{subfigure}[b]{0.32\textwidth}
 \centering
 \includegraphics[width=\textwidth]{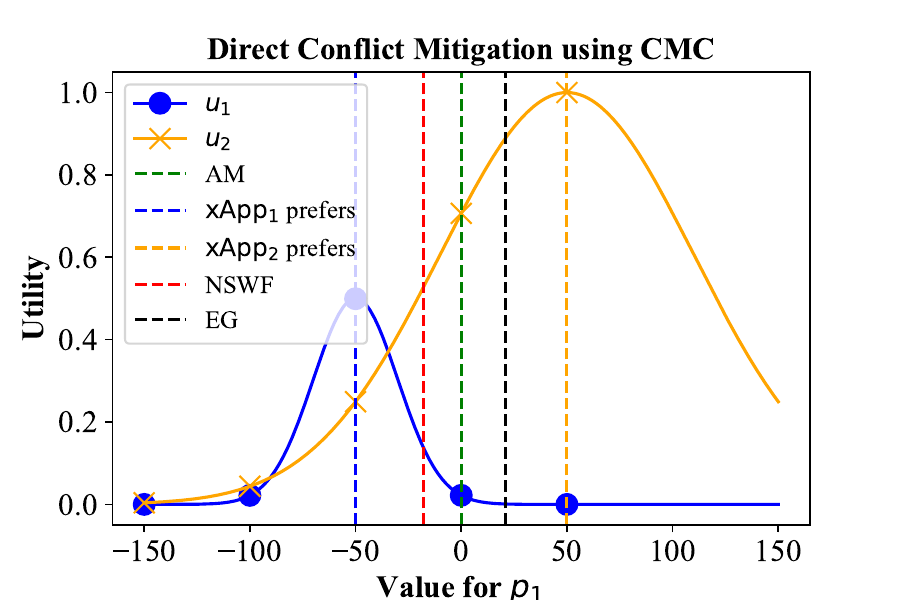}
 \caption{Direct conflict ($\rm{xApp_1~ and ~xApp_2}$)}
 \label{fig:res_conf_a}
  \end{subfigure}
  \hfill
  \begin{subfigure}[b]{0.32\textwidth}
 \centering
 \includegraphics[width=\textwidth]{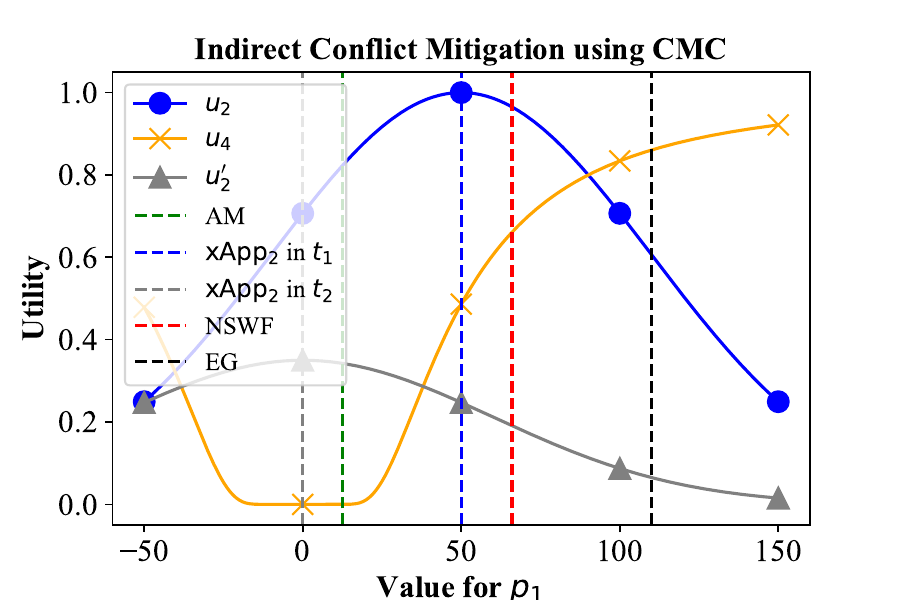}
 \caption{Indirect conflict ($\rm{xApp_2~ and ~xApp_4}$)}
 \label{fig:res_conf_b}
  \end{subfigure}
  \hfill
  \begin{subfigure}[b]{0.32\textwidth}
 \centering
 \includegraphics[width=\textwidth]{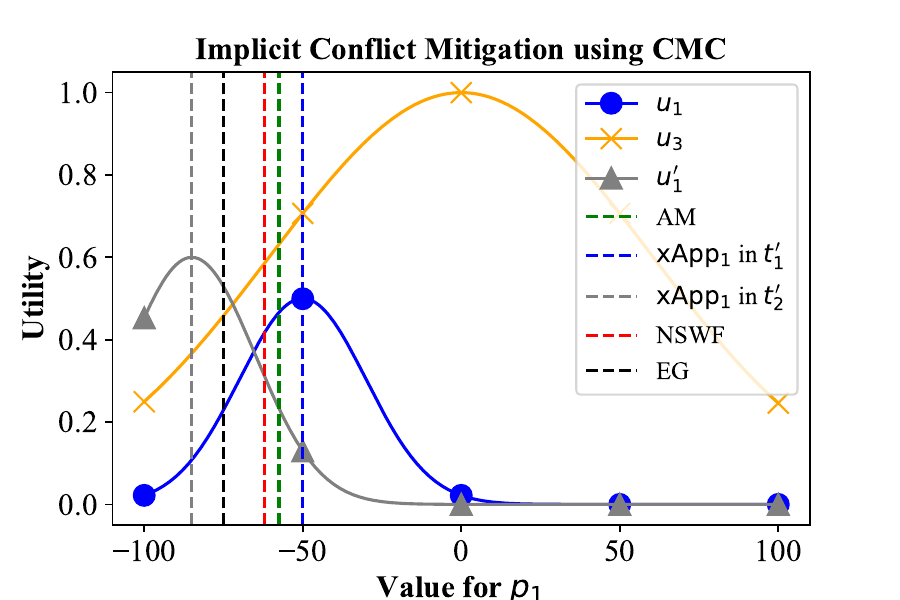}
 \caption{Implicit conflict ($\rm{xApp_1~ and ~xApp_3}$)}
 \label{fig:res_conf_c}
  \end{subfigure}
  \caption{Conflict mitigation using the proposed \ac{CMC}.}
  \label{fig:res_conf}
  \vspace{-0.3in}
\end{figure*}

\section{Numerical Results and Analysis}\label{sec:num_res}
To assess the performance of the proposed \ac{CMC}, we construct an example model incorporating direct, indirect, and implicit conflicts, as illustrated in Fig.~\ref{fig:exam_model}. During the numerical analysis, we assume that the output of all \acp{xApp}, representing the \acp{KPI}, follows a Gaussian distribution function, as it often mirrors real-life scenarios \cite{b9}. The equations ($o_1 = 0.5 \times e^{-\frac{(p_1 +50)^2}{2p_2^2}}$, $o_2 = e^{-\frac{(p_1 -50)^2}{2p_3^2}}$, $o_3 = e^{-\frac{(p_4 + o_1)^2}{2p_5^2}}$, $o_4 = e^{-\frac{(p_7 + o_2)^2}{2p_6^2}}$, $o_1^\prime = 0.6 \times e^{-\frac{(p_1 - 85)^2}{2p_2^2}}$, $o_2^\prime = 0.35 \times e^{-\frac{(p_1 + 0)^2}{2p_3^2}}$) and parameters are configured to accurately represent the mentioned conflicts. Indirect and implicit conflicts are modeled similarly, as it is challenging to precisely model implicit conflict, and both types can be mitigated using a similar approach.

\par All obtained \acp{KPI} from the \acp{xApp} in Fig.~\ref{fig:exam_model} are converted to utilities within the range $[0, 1]$ using the formula stated in section~\ref{theory_assump}. Priority weights assigned by the \ac{MNO} to \acp{xApp} for direct, indirect, and implicit conflicts are assumed to be $\{\rm{xApp_1} = 0.4~ \rm{and} ~\rm{xApp_2} = 0.6\}$, $\{\rm{xApp_2} = 0.1~ \rm{and} ~\rm{xApp_4} = 0.9\}$, and $\{\rm{xApp_1} = 0.9~ \rm{and} ~\rm{xApp_3} = 0.1\}$, respectively. These weights are utilized only for the \ac{EG} solution when the \ac{MNO} prioritizes an \ac{xApp} over others based on the network state and demands. The \ac{NSWF} and \ac{EG} solutions are effective in non-priority and priority scenarios, respectively. We perform \ac{AM} with different values for the conflicting parameter and obtain the optimal value for comparison with the performance of \ac{NSWF} in a non-priority scenario. Each intelligent \acp{xApp} is trained using Polynomial Regression Blocks in Python, leveraging the generated dataset available in \cite{b11} through the aforementioned equations.

In Fig.~\ref{fig:res_conf}(a), we illustrated the direct conflict mitigation between $\rm{xApp}_1$ and $\rm{xApp}_2$ over the parameter $p_1$. Based on the current network state assumed in the example model, $\rm{xApp}_1$ demands $p_1 = 50$ to obtain its objective, whereas $\rm{xApp}_2$ demands $p_1 = -50$; this creates a direct conflict between these two \acp{xApp}. To resolve this conflict, we used \ac{NSWF}, \ac{AM}, and \ac{EG} solutions to suggest a value for $p_1$ as demonstrated in Fig.~\ref{fig:res_conf}(a). The \ac{NSWF} ensures maximum collective utility over \ac{AM} and \ac{EG}, whereas \ac{EG} prioritizes $\rm{xApp}_2$'s utility over $\rm{xApp}_1$'s as per the priority weight assigned by the \ac{MNO}.

To demonstrate the capability of the proposed \ac{CMS} in handling indirect conflicts, we focus on $\rm{xApp}_2$ and $\rm{xApp}_4$. The utility functions of these \acp{xApp} are designed to have a logical dependency, leading to an indirect conflict scenario. When $\rm{xApp}_2$ modifies parameter $p_1$ at timestamp $t_2$ from its previous value at $t_1$, it indirectly impacts the scope of $\rm{xApp}_4$, resulting in a reduction of its utility as illustrated in Fig.~\ref{fig:res_conf}(b). The change made by $\rm{xApp}_2$ at $t_2$, setting $p_1 = 0$ based on the network state, indirectly affects the utility of $\rm{xApp}_4$, nearly reducing it to zero. 

\par This resembles the impact of the energy-saving \ac{xApp} altering transmission power and indirectly influencing the handover boundary, thereby reducing the utilities associated with mobility robustness optimization \ac{xApp}. Comparing the optimal values suggested by \ac{AM} and \ac{NSWF}, we observe that \ac{AM} does not improve the utility of $\rm{xApp}_4$, while \ac{NSWF} significantly enhances the collective utility between the two conflicting \acp{xApp} (see Fig.~\ref{fig:res_conf}(b)). Additionally, the \ac{EG} solution verifies higher utility for $\rm{xApp}_4$ when it is prioritized by the \ac{MNO}. 


\par To analyze the implicit conflict between $\rm{xApp}_1$ and $\rm{xApp}_3$, we formulate the utility equation for $\rm{xApp}_3$ to capture any inadvertent or unwanted impact, similar to an indirect conflict resulting from $\rm{xApp}_1$ changing $p_1$. As depicted in Fig.~\ref{fig:res_conf}(c), the implicit conflict arises when $\rm{xApp}_1$ sets $p_1 = -85$ at timestamp $t_2^\prime$ from its value at $t_1^\prime$, leading to a reduction in the utility of $\rm{xApp}_3$ below $0.4$, causing dissatisfaction and triggering the \ac{CMC}. Subsequently, the \ac{CMC} suggests optimal values for $p_1$ using \ac{AM}, \ac{NSWF}, and \ac{EG}. Both \ac{AM} and \ac{NSWF} improve the utility for $\rm{xApp}_3$ while reducing that of $\rm{xApp}_1$. On the other hand, the \ac{EG} solution enhances the utility for $\rm{xApp}_1$ when it is prioritized by the Mobile Network Operator (MNO), establishing a balance between the utilities of $\rm{xApp}_1$ and $\rm{xApp}_3$ at $t_2^\prime$.

\par The indirect conflict occurs when $\rm{xApp}_2$ requests a different setting of $p_1$ at timestamp $t_2$ while the $\rm{xApp}_4$ remains with the same setting. Therefore, we only illustrated changes in $\rm{xApp}_2$'s utilities at different timestamps in Fig.~\ref{fig:res_conf}(b). Similar explanation is applied to $\rm{xApp}_1$ and $\rm{xApp}_3$ in Fig.~\ref{fig:res_conf}(c).


\section{Limitations and Challenges}
\label{sec:limit}
After critically analyzing the current state of the proposed conflict mitigation system, we can underline the following limitations and challenges of the proposed approach: 

\begin{itemize}
 \item A Higher number of \acp{xApp} and \acp{ICP} involved in a conflicting scenario may increase complexity to reach the optimal equilibrium state and hence may exceed the time threshold of \ac{Near-RT-RIC} functions. 
 
 \item Converting \acp{KPI} to utilities can be a challenge in a real-world setting. 

 \item The proposed solution is only tested in a theory-based example scenario which does not assure its applicability in a real-world system. 

 \item The deployment of the proposed solution has a specific infrastructure prerequisite to be fulfilled. 

 \item The benefit of the proposed approach can only become conclusive after its successful deployment in a real-world setting with real \acp{xApp}. 
 
\end{itemize}

\section{Conclusion and Future Works}\label{sec:con_fut}
In this article, we have presented a novel game-theory based \ac{CMS} for the \ac{Near-RT-RIC} to ensure an intelligent and vendor-agnostic Open \ac{RAN} architecture. Our proposed system is designed to address the issue of conflicts arising from the coexistence of multiple \acp{xApp} from different vendors. The core contribution of our work lies in the establishment of an independent conflict detection and mitigation controller to mitigate conflicts. By introducing a game-theory-based independent conflict mitigation controller, we demonstrate effective mitigation of conflicts between \acp{xApp} in an experimental setup. 

\par The application of cooperative bargain game theory, specifically the \ac{NSWF} and the \ac{EG} solution, allowed us to determine optimal configurations for conflicting parameters, ensuring maximum collective utility for the network. Moving forward, we intend to transition from the theoretical model to a more practical real-world setup and overcome the current limitations of the proposed approach. We plan to reduce the complexity and computational overhead for scenarios with a high density of \acp{xApp} and \acp{ICP}. Additionally, we plan to explore machine learning techniques to predict probable conflicts and multi-agent reinforcement learning techniques for mitigating the detected conflicts.

\section*{Acknowledgment}
We acknowledge funding from the European Union's Horizon Europe research and innovation program under the Marie Skłodowska-Curie SE grant agreement RE-ROUTE No 101086343.


\end{document}

%% file: Acronyms.tex
\begin{acronym}[BEREC]
\acro{10G-EPON}{10 Gigabit EPON}
\acro{3GPP}{3rd Generation Partnership Project}
\acro{5G}{Fifth Generation}
\acro{6G}{Sixth Generation}
\acro{A-CPI}{A Controller Plane Interface}
\acro{A-RoF}{Analog Radio-over-Fiber}
\acro{ABNO}{Application-Based Network Operations}
\acro{ADC}{Analog-to-Digital Converter}
\acro{AE}{Allocative Efficiency}
\acro{amc}{adaptive modulation and coding}
\acro{AON}{Active Optical Network}
\acro{AI}{Artificial Intelligence}
\acro{ML}{Machine Learning}
\acro{API}{Application Programming Interface}
\acro{AWG}{Arrayed Waveguide Grating}
\acro{B2B}{Business to Business}
\acro{B2C}{Business to Consumer}
\acro{B-RAS}{Broadband Remote Access Servers}
\acro{BaaS}{Blockchain as a Service}
\acro{BB}{Budget Balance}
\acro{BBF}{BroadBand Forum}
\acro{BBU}{Base Band Unit}
\acro{BER}{Bit Error Rate}
\acro{BEREC}{Body of European Regulators for Electronic Communications}
\acro{BF}{beamforming}
\acro{BFT}{Byzentine fault tolerant}
\acro{BGP}{Border Gateway Protocol}
\acro{BMap}{Bandwidth Map}
\acro{BPM}{Business Process Management}
\acro{BS}{Base Station}
\acro{BSC}{Base Station Controller}
\acro{BtB}{Back to Back}
\acro{BW}{Bandwidth}
\acro{C2C}{Consumer to Consumer}
\acro{C-RAN}{Cloud Radio Access Network}
\acro{C-RoFN}{Cloud-based Radio over optical Fiber Network}
\acro{CA}{Certificate Authority}
\acro{CapEx}{Capital Expenditure}
\acro{CDMA}{code division multiple access}
\acro{CDR}{Clock Data Recovery}
\acro{CFO}{carrier frequency offset}
\acro{CIR}{Committed Information Rate}
\acro{CO}{Central Office}
\acro{CoMP}{Coordinated Multipoint}
\acro{COP}{Control Orchestration Protocol}
\acro{CORD}{Central Office Rearchitected as a Data Centre}
\acro{CPE}{Customer Premises Equipment}
\acro{CPRI}{Common Public Radio Interface}
\acro{CPU}{Central Processing Unit}
\acro{CMC}{Conflict Mitigation Controller}
\acro{CMS}{Conflict Management System}
\acro{CQI}{channel quality indicator}
\acro{CR}{cognitive radio}
\acro{CS}{Central Station}
\acro{CSI}{channel state information}
\acro{D2D}{Device to Device}
\acro{D-CPI}{D Controller Plane Interface}
\acro{D-RoF}{Digital Radio-over-Fiber}
\acro{DAC}{Digital-to-Analog Converter}
\acro{DAS}{distributed antenna system}
\acro{DAS}{Distributed Antenna Systems}
\acro{DBA}{Dynamic Bandwidth Allocation}
\acro{DBRu}{Dynamic Bandwidth Report upstream}
\acro{DC}{Direct Current}
\acro{DL}{Downlink}
\acro{DLT}{Distributed Ledger Technology}
\acro{DMT}{Discrete Multitone}
\acro{DNS}{Domain Name Server}
\acro{DPDK}{Data Plane Development Kit}
\acro{DSB}{Double Side Band}
\acro{DSL}{Digital Subscriber Line}
\acro{DSLAM}{Digital Subscriber Line Access Multiplexer}
\acro{DSP}{Digital Signal Processing}
\acro{DSS}{Distributed Synchronization Service}
\acro{DU}{Distributed Unit}
\acro{DUE}{D2D user equipment}
\acro{DWDM}{Dense Wave Division Multiplexing}
\acro{E-CORD}{Enterprise CORD}
\acro{EAL}{Environment Abstraction Layer}
\acro{EDF}{Erbium-Doped Fiber}
\acro{EFM}{Ethernet in the First Mile}
\acro{EMBS}{Elastic Mobile Broadband Service}
\acro{EON}{Elastic Optical Network}
\acro{EPC}{Evolved Packet Core}
\acro{EPON}{Ethernet Passive Optical Network}
\acro{eMBB}{enhanced Mobile Broadband}
\acro{ETSI}{European Telecommunications Standards Institute}
\acro{EVM}{Error Vector Magnitude}
\acro{FANS}{Fixed Access Network Sharing}
\acro{FASA}{Flexible Access System Architecture}
\acro{FBMC}{filter bank multi-carrier}
\acro{FBMC}{Filterbank Multicarrier}
\acro{FCC}{Federal Communications Commission}
\acro{FDD}{Frequency Division Duplex}
\acro{FDMA}{frequency-division multiple access}
\acro{FEC}{Forward Error Correction}
\acro{FFR}{Fractional Frequency Reuse}
\acro{FFT}{Fast Fourier Transform}
\acro{FPGA}{Field-Programmable Gate Array}
\acro{FSO}{Free-Space-Optics}
\acro{FSR}{Free Spectral Range}
\acro{FTN}{faster-than-nyquist}
\acro{FTTC}{Fiber-to-the-Curb}
\acro{FTTH}{Fiber-to-the-Home}
\acro{FTTx}{Fiber-to-the-x}
\acro{G.Fast}{Fast Access to Subscriber Terminals}
\acro{GEM}{G-PON encapsulation method}
\acro{GFDM}{Generalized Frequency Division Multiplexing}
\acro{GMPLS}{Generalized Multi-Protocol Label Switching}
\acro{GPON}{Gigabit Passive Optical Network}
\acro{GPP}{General Purpose Processor}
\acro{GPRS}{General Packet Radio Service}
\acro{GSM}{Global System for Mobile Communications}
\acro{GTP}{GPRS Tunneling Protocol}
\acro{GUI}{Graphical User Interface}
\acro{HA}{Hardware Accelerator}
\acro{HARQ}{Hybrid-Automatic Repeat Request}
\acro{HD}{half duplex}
\acro{HetNet}{heterogeneous networks}
\acro{IaaS}{Infrastructure as a Service}
\acro{I-CPI}{I Controller Plane Interface}
\acro{I2RS}{Interface 2 Routing System}
\acro{IA}{Interference Alignment}
\acro{IAM}{Identity and Access Management}
\acro{IBFD}{in-band full duplex}
\acro{IC}{Incentive Compatibility}
\acro{ICIC}{inter-cell interference coordination}
\acro{ICT}{information and communications technology}
\acro{IEEE}{Institute of Electrical and Electronics Engineers}
\acro{IETF}{Internet Engineering Task Force}
\acro{IF}{Intermediate Frequency}
\acro{IFFT}{inverse FFT}
\acro{ITS}{Intelligent Transport Systems}
\acro{InP}{Infrastructure Provider}
\acro{IoT}{Internet of Things}
\acro{IP}{Internet Protocol}
\acro{IQ}{In-phase/Quadrature}
\acro{IR}{Individual Rationality}
\acro{IRC}{Interference Rejection Combining}
\acro{ISI}{inter-symbol interference}
\acro{ISP}{Internet Service Provider}
\acro{BFT}{Byzantine-Fault-Tolerant}
\acro{IV}{Intelligent Vehicle}
\acro{ITU}{International Telecommunication Union}
\acro{KPI}{Key Performance Indicator}
\acro{KVM}{Kernel-based Virtual Machine}
\acro{L2}{Layer-2}
\acro{L3}{Layer-3}
\acro{LAN}{Local Area Network}
\acro{LO}{Local Oscillator}
\acro{LOS}{Line Of Sight}
\acro{LR-PON}{Long Reach PON}
\acro{LSP-DB}{Label Switched Path Database}
\acro{LTE-A}{Long Term Evolution Advanced}
\acro{LTE}{Long Term Evolution}
\acro{M-CORD}{Mobile CORD}
\acro{M-MIMO}{massive MIMO}
\acro{M2M}{machine-to-machine}
\acro{MAC}{Medium Access Control}
\acro{ME}{Merging Engine}
\acro{MIMO}{multiple input multiple output}
\acro{MME}{Mobility Management Entity}
\acro{mmWave}{millimeter wave}
\acro{MNO}{mobile network operator}
\acro{MPLS}{Multiprotocol Label Switching}
\acro{MRC}{Maximum Ratio Combining}
\acro{MSC}{Mobile Switching Centre}
\acro{MSP}{Membership Service Providers}
\acro{MSR}{Multi-Stratum Resources}
\acro{MTC}{machine type communication}
\acro{mMTC}{massive Machine Type Communications}
\acro{MVNO}{Mobile Virtual Network Operator}
\acro{MVNP}{mobile virtual network provider}
\acro{MZI}{Mach-Zehnder Interferometer}
\acro{MZM}{Mach-Zehnder Modulator}
\acro{NE}{Network Element}
\acro{NFV}{Network Function Virtualization}
\acro{NFVaaS}{Network Function Virtualization as a Service}
\acro{NG-PON2}{Next-Generation Passive Optical Network 2}
\acro{NGA}{Next Generation Access}
\acro{NLOS}{None Line Of Sight}
\acro{NMS}{Network Management System}
\acro{NRZ}{Non Return-to-Zero}
\acro{NTT}{Nippon Telegraph and Telephone}
\acro{OBPF}{Optical BandPass Filter}
\acro{OBSAI}{Open Base Station Architecture Initiative}
\acro{ODN}{Optical Distribution Network}
\acro{OFC}{Optical Frequency Comb}
\acro{OFDM}{orthogonal frequency-division multiplexing}
\acro{OFDMA}{orthogonal frequency-division multiple access}
\acro{OLO}{Other Licensed Operator}
\acro{OLT}{Optical Line Terminal}
\acro{ONF}{Open Networking Foundation}
\acro{ONU}{Optical Network Unit}
\acro{OOB}{out-of-band}
\acro{OOK}{On-off Keying}
\acro{OpenCord}{Central Office Re-Architected as a Data Center}
\acro{OpEx}{Operating Expenditure}
\acro{OSI}{Open Systems Interconnection}
\acro{OSS}{Operations Support Systems}
\acro{OTT}{Over-the-Top}
\acro{OXM}{OpenFlow Extensible Match}
\acro{P2MP}{Ethernet over point-to-multipoint }
\acro{P2P}{Point-to-Point }
\acro{PAM}{Pulse Amplitude Modulation}
\acro{PAPR}{peak-to-average power rating}
\acro{PAPR}{Peak-to-Average Power Ratio}
\acro{PBMA}{Priority Based Merging Algorithm}
\acro{PCE}{Path Computation Elements}
\acro{PCEP}{Path Computation Element Protocol}
\acro{PCF}{Photonic Crystal Fiber}
\acro{PD}{Photodiode}
\acro{PDCP}{RD Control Protocol}
\acro{PDF}{probability distribution function}
\acro{PGW}{Packet Gateway}
\acro{PHY}{physical Layer}
\acro{PIR}{Peak Information Rate}
\acro{PMD}{Polarization Division Multiplexing}
\acro{PON}{Passive Optical Network}
\acro{POTS}{Plain Old Telephone Service}
\acro{PoET}{Proof of Elapsed Time}
\acro{PoW}{Proof of Work}
\acro{PoS}{Proof of Stake}
\acro{PTP}{Precision Time Protocol}
\acro{PWM}{Pulse Width Modulation}
\acro{QAM}{Quadrature Amplitude Modulation}
\acro{QoE}{Quality of Experience}
\acro{QoS}{Quality of Service}
\acro{QPSK}{Quadrature Phase Shift Keying}
\acro{R-CORD}{Residential CORD}
\acro{RA}{Resource Allocation}
\acro{RAN}{Radio Access Network}
\acro{RAT}{Radio Access Technology}
\acro{RB}{resource block}
\acro{RFIC}{radio frequency integrated circuit}
\acro{RIC}{RAN Intelligent Controller}
\acro{RLC}{Radio Link Control}
\acro{RN}{Remote Node}
\acro{ROADM}{Reconfigurable Optical Add Drop Multiplexer}
\acro{RoF}{Radio-over-Fiber}
\acro{RRC}{Radio Resource Control}
\acro{RRH}{Remote Radio Head}
\acro{RRPH}{Remote Radio and PHY Head}
\acro{RRU}{Remote Radio Unit}
\acro{RSOA}{Reflective Semiconductor Optical Amplifier}
\acro{RU}{Remote Unit}
\acro{Rx}{receiver}
\acro{SC-FDMA}{single carrier frequency division multiple access}
\acro{SCM}{single carrier modulation}
\acro{SD-RAN}{Software Defined Radio Access Network}
\acro{SDAN}{Software Defined Access Network}
\acro{SDMA}{Semi-Distributed Mobility Anchoring}
\acro{SDN}{Software Defined Networking}
\acro{SDR}{Software Defined Radio}
\acro{SFBD}{Single Fiber Bi-Direction}
\acro{SGW}{Serving Gateway}
\acro{SI}{self-interference}
\acro{SIC}{self-interference cancellation}
\acro{SIMO}{Single Input Multiple Output}
\acro{SLA}{Service Level Agreement}
\acro{SMF}{Single Mode Fiber}
\acro{SNMP}{Simple Network Management Protocol}
\acro{SNR}{Signal-to-Noise Ratio}
\acro{SP}{Service Providers}
\acro{Split-PHY}{Split Physical Layer}
\acro{SRI-OV}{Single Root Input/Output Virtualization}
\acro{SU}{secondary user}
\acro{SUT}{System Under Test}
\acro{T-CONT}{Transmission Container}
\acro{TC}{Transmission Convergence}
\acro{TCO}{Total Cost of Ownership}
\acro{TD-LTE}{Time Division LTE}
\acro{TDD}{Time Division Duplex}
\acro{TDM}{Time Division Multiplexing}
\acro{TDMA}{Time Division Multiple Access}
\acro{TED}{Traffic Engineering Database}
\acro{TEID}{Tunnel endpoint identifier}
\acro{TLS}{Transport Layer Security}
\acro{TPS}{Transactions per Second}
\acro{TTI}{transmission time interval}
\acro{TWDM}{Time and Wavelength Division Multiplexing}
\acro{Tx}{transmitter}
\acro{UD-CRAN}{Ultra-Dense Cloud Radio Access Network}
\acro{UDP}{User Datagram Protocol}
\acro{UE}{User Equipment}
\acro{UF-OFDM}{Universally Filtered OFDM}
\acro{UFMC}{Universally Filtered Multicarrier}
\acro{UL}{Uplink}
\acro{UMTS}{Universal Mobile Telecommunications System}
\acro{URLLC}{Ultra-Reliable Low-Latency Communication}
\acro{USRP}{Universal Software Radio Peripheral}
\acro{V2I}{Vehicle to Infrastructure}
\acro{V2V}{Vehicle to Vehicle}
\acro{vBBU}{virtualized BBU}
\acro{vBMap}{Virtual Bandwidth Map}
\acro{vBS}{virtual Base station}
\acro{VCG}{Vickery-Clarke-Groves}
\acro{vCPE}{virtual CPE}
\acro{vCPU}{Virtual Central Processing Unit}
\acro{vDBA}{virtual DBA}
\acro{VDSL2}{Very-high-bit-rate digital subscriber line 2}
\acro{VLAN}{Virtual Local Area Network}
\acro{VM}{Virtual Machine}
\acro{VNF}{Virtual Network Functions}
\acro{VNI}{visual networking index}
\acro{VNO}{Virtual Network Operator}
\acro{VNTM}{Virtual Network Topology Manager}
\acro{vOLT}{virtual OLT}
\acro{VPE}{virtual Provider Edge}
\acro{VPN}{Virtual Private Network}
\acro{V2X}{Vehicle to Everything}
\acro{VTN}{Virtual Tenant Network}
\acro{VULA}{Virtual Unblunded Local Access}
\acro{WAN}{Wide Area Network}
\acro{WAP}{Wireless Access Point}
\acro{WCDMA}{wide-band code division multiple access}
\acro{WDM-PON}{Wavelength Division Multiplexing - Passive Optical Network}
\acro{WDM}{Wavelength Division Multiplexing}
\acro{WiMAX}{Worldwide Interoperability for Microwave Access}
\acro{WRPR}{Wired-to-RF Power Ratio}
\acro{XG-PON}{10 Gigabit PON}
\acro{XGEM}{XG-PON encapsulation method}
\acro{XOS}{XaaS Operating System}

\acro{RIC}{RAN Intelligent Controller}
\acro{Near-RT-RIC}{Near Real Time RAN Intelligent Controller}
\acro{Non-RT-RIC}{Non Real Time RAN Intelligent Controller}
\acro{xApp}{Extended Application}
\acro{rApp}{Remote Application}
\acro{CMS}{Conflict Mitigation System}
\acro{EG}{Eisenberg-Galle}
\acro{NSWF}{Nash's Social Welfare Function} 
\acro{CMC}{Conflict Mitigation Controller}
\acro{CDC}{Conflict Detection Controller}
\acro{PMon}{Performance Monitoring}
\acro{RCP}{Recently Changed Parameter}
\acro{PGD}{Parameter Group Definition}
\acro{RCPG}{Recently Changed Parameter Group}
\acro{DCKD}{Decision Correlated with KPI Degradation} 
\acro{KDO}{KPI Degradation Occurrences}
\acro{MNO}{Mobile Network Operator}
\acro{SDL}{Shared Data Layer}
\acro{SON}{Self Organising Network}
\acro{SF}{SON Function}
\acro{AM}{Arithmetic Mean}
\acro{PKR}{Parameter and KPI Ranges}
\acro{KPI}{Key Performance Indicator}
\acro{ICP}{Input Control Parameter}
\acro{RF}{Radio Frequency}

\end{acronym}

%% file: manuscript.bbl
\vspace{-10pt}

\begin{thebibliography}{00}
\bibitem{b1} Singh, Sameer Kumar, Rohit Singh, and Brijesh Kumbhani. "The evolution of radio access network towards open-RAN: Challenges and opportunities.", (WCNCW), pp. 1-6. IEEE, 2020.

\bibitem{b2} Polese, Michele, Leonardo Bonati, Salvatore D’Oro, Stefano Basagni, and Tommaso Melodia. "Understanding O-RAN: Architecture, interfaces, algorithms, security, and research challenges." IEEE Communications Surveys \& Tutorials (2023).

\bibitem{rimedo_labs} Dryjański, Marcin. "Traffic Management for V2X Use Cases in O-RAN." Available online: \url{https://rimedolabs.com/blog/traffic-management-for-v2x-use-cases-in-o-ran/} [Accessed on: 09-11-2023].

\bibitem{b3} O-RAN Alliance, O-RAN Use Cases Analysis Report, O-RAN Al- liance, October 2022, v09.00.
\bibitem{b4} O-RAN Working Group 3, Near-Real-time RAN Intelli- gent Controller Architecture, O-RAN Alliance, October 2022, v03.00.
\bibitem{b5} O-RAN Working Group 1, O-RAN Architecture Description, O-RAN Alliance, October 2022, v07.00.
\bibitem{b6} Adamczyk, Cezary, and Adrian Kliks. "Conflict Mitigation Framework and Conflict Detection in O-RAN Near-RT RIC." IEEE Communications Magazine (2023).
\bibitem{b7} Zhang, Han, Hao Zhou, and Melike Erol-Kantarci. "Team learning-based resource allocation for open radio access network (O-RAN)." In ICC 2022-IEEE International Conference on Communications, pp. 4938-4943. IEEE, 2022.

\bibitem{b8}Ramezani, Sara, and Ulle Endriss. "Nash social welfare in multiagent resource allocation." In International Workshop on Agent-Mediated Electronic Commerce, pp. 117-131. Berlin, Heidelberg: Springer Berlin Heidelberg, 2009.

\bibitem{b9} Banerjee, Anubhab, Stephen S. Mwanje, and Georg Carle. "Toward control and coordination in cognitive autonomous networks." IEEE Transactions on Network and Service Management 19, no. 1 (2021): 49-60.

\bibitem{b10} Brânzei, Simina, Vasilis Gkatzelis, and Ruta Mehta. "Nash social welfare approximation for strategic agents." In Proceedings of the 2017 ACM Conference on Economics and Computation, pp. 611-628. 2017.

\bibitem{b11} \url{https://github.com/dewanwadud1/cmsORAN}.

\end{thebibliography}
